\documentclass[a4paper]{article}

\usepackage{INTERSPEECH2016}

\usepackage{amsmath,graphicx}
\usepackage{amssymb,amsmath,bm}
\usepackage{textcomp}
\usepackage{algorithm}
\usepackage{algpseudocode}
\usepackage{float}
\usepackage{multirow}
\usepackage[justification=centering]{caption}
\usepackage[caption = false]{subfig}

\usepackage{tikz,placeins}
\usetikzlibrary{arrows,decorations.markings}
\tikzstyle{every picture}+=[font=\rmfamily\it\bfseries\large]
\usetikzlibrary{positioning, shapes}
\pgfdeclarelayer{background}
\pgfdeclarelayer{foreground}
\pgfsetlayers{background,main,foreground}

\graphicspath{{.}{../figures/}}
\makeatletter
  \def\input@path{{.}{../figures/}}
\makeatother

\newcommand{\specialcell}[2][c]{%
     \begin{tabular}[#1]{@{}c@{}}#2\end{tabular}}

\ninept

\title{Efficient Implementation of the Room Simulator \\ for Training Deep Neural Network Acoustic Models}
\name{{Chanwoo Kim$^{1\dagger}$\thanks{$^\dagger$Work performed while at Google.}, Ehsan Variani$^2$, Arun Narayanan$^2$, and Michiel Bacchiani$^2$}}

\address{$^1$Samsung Research, $^2$Google Speech \\
  {\small \tt $^1$chanw.com@samsung.com $^2$\{variani, arunnt, michiel\}@google.com }}
\begin{document}
\maketitle
\begin{abstract}

In this paper, we describe how to efficiently implement
an acoustic room simulator to generate large-scale
simulated data for training deep neural networks.
Even though \textit{Google Room Simulator} in \cite{C_Kim_INTERSPEECH_2017_1} was shown
 to be quite effective in reducing the Word
Error Rates (WERs) for far-field applications by generating
simulated far-field training sets, it requires a very large 
number of FFTs.
\textit{Room Simulator} used approximately 80 \% of CPU usage
  in our \textit{CPU/GPU} training architecture
  \cite{E_Variani_INTERSPEECH_2017_01}.
  In this work, we implement an efficient OverLap
  Addition (OLA) based filtering using the open-source \texttt{FFTW3}
library. Further, we investigate the effects of the Room
Impulse Response (RIR) lengths. Experimentally, we conclude that we can cut
the tail portions of RIRs whose power is less than 20 \textit{dB}
below the maximum power without sacrificing the speech recognition accuracy.
However, we observe that cutting  RIR tail more than this threshold
harms the speech recognition accuracy for rerecorded test sets.
Using these approaches, we were able to reduce CPU usage for the
room simulator portion down to 9.69 \%
in CPU/GPU training architecture. Profiling result shows that
we obtain 22.4 times speed-up on a single machine and 37.3 times
  speed up on Google's distributed training infrastructure.
 \end{abstract}
%
\section{Introduction}
With advancements in deep learning
\cite{Seltzer2013DNNAurora4, Yu2013FeatureLearningDNN, V_Vanhoucke_Deep_Learning_NIPS_Workshop_2011,
G_Hinton_IEEE_Signal_Process_Mag_2012,
T_Sainath_IEEETran_2017_1, T_Sainath_Book_Chapter_2017_1},
speech recognition accuracy has improved dramatically.
Now, speech recognition systems
are used not only on portable devices
but also on standalone devices for far-field speech recognition.
Examples include voice assistant systems such as Amazon Alexa
and Google Home \cite{C_Kim_INTERSPEECH_2017_1, B_Li_INTERSPEECH_2017_1}.
In far-field speech recognition, the impact of noise and reverberation
is much larger than near-field cases. Traditional approaches to far-field
speech recognition include noise robust feature extraction algorithms
\cite{C_Kim_IEEETran_2016_1, U_H_Yapanel_SpeechComm_2008, C_Kim_ICASSP_2010_1},
on-set enhancement algorithms
\cite{C_Kim_INTERSPEECH_2010_2, C_Kim_INTERSPEECH_2014_2}, and multi-microphone
approaches \cite{T_Nekatani_ICASSP_2017_1, T_Higuchi_ICASSP_2016_1,
H_Erdogan_INTERSPEECH_2016_1, C_Kim_INTERSPEECH_2015_1, C_Kim_INTERSPEECH_2009_1, C_Kim_ICASSP_2012_2, C_Kim_ICASSP_2011_2}.
Recently, we observed that training with large-scale noisy data generated
by a \textit{Room Simulator} \cite{C_Kim_INTERSPEECH_2017_1}
improves speech recognition accuracy dramatically.
This system has been successfully employed
for training acoustic models for Google Home or Google voice
search \cite{C_Kim_INTERSPEECH_2017_1}.

\begin{figure}
  \begin{center}
    \resizebox{75mm}{!}{\input{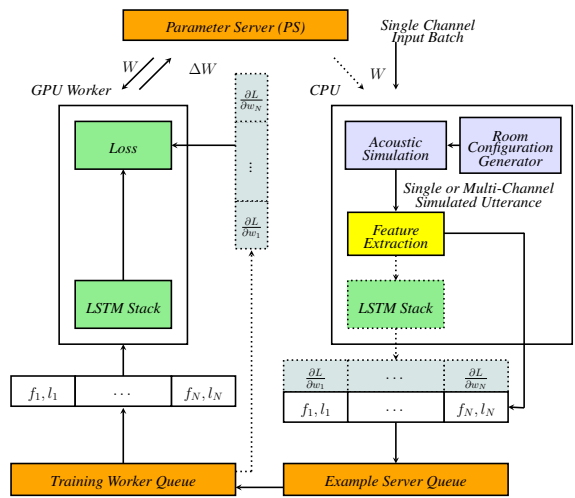}}
      \caption {
        \label{fig:gpu_cpu_structure}
        Training architecture using cluster of CPUs and GPUs.
        \cite{E_Variani_INTERSPEECH_2017_01}.
        \textit{Room Simulator} is on the CPU side to generate
        simulated utterances using room configurations.
        \cite{C_Kim_INTERSPEECH_2017_1}.}
  \end{center}
\vspace{-10mm}
\end{figure}

\textit{Room Simulator} creates millions of virtual rooms
with different dimensions and different number of sound sources
at different locations and Signal-to-Noise Ratios (SNRs).
For every new utterance in the training set, we use a randomly
sampled room configuration, so that the same utterance
is simulated under different acoustic environments in every epoch during training.
As will be seen in Sec. \ref{sec:experimental_results}, if
we generate the simulated utterance only once for each
input example, the performance is worse.
Since different RIRs are applied for the same
utterance at every epoch, the intermediate results cannot be cached.
Therefore, the noisification process requires very large number of convolution operations. As will be seen
in Sec. \ref{sec:experimental_results}, during training, \textit{Room Simulator} used up to 80 \%
of the whole CPU usage if we use the CPU/GPU architecture
shown in Fig. \ref{fig:gpu_cpu_structure}.
In this paper, we describe our approach to reduce the
computational portion of the \textit{Room Simulator} below 10 \% of the CPU usage
in our CPU/GPU training scheme.
%
%
%
%
%
\section{Efficient Implementation of the Room Simulation System}
  \label{sec:implementation}
Fig. \ref{fig:gpu_cpu_structure} shows a high level block diagram of
the acoustic model training infrastructure
\cite{E_Variani_INTERSPEECH_2017_01}. Since it is not easy to
run all the front-end processing blocks such as feature
extraction and Voice Activity Detection (VAD) on GPUs,
 they run on CPUs during training.
Even though Fast Fourier Transform (FFT) of the
\textit{Room Simulator} may be efficiently implemented on GPUs,
since
\textit{Room Simulator} must feed the simulated utterances
to the rest of CPU optimized front-end components, the Room Simulator
is also running on CPUs.
%
%
%
%
\subsection{Review of the Room Simulator for Data Augmentation}
\begin{figure}
  \begin{center}
    \includegraphics[width=70mm]{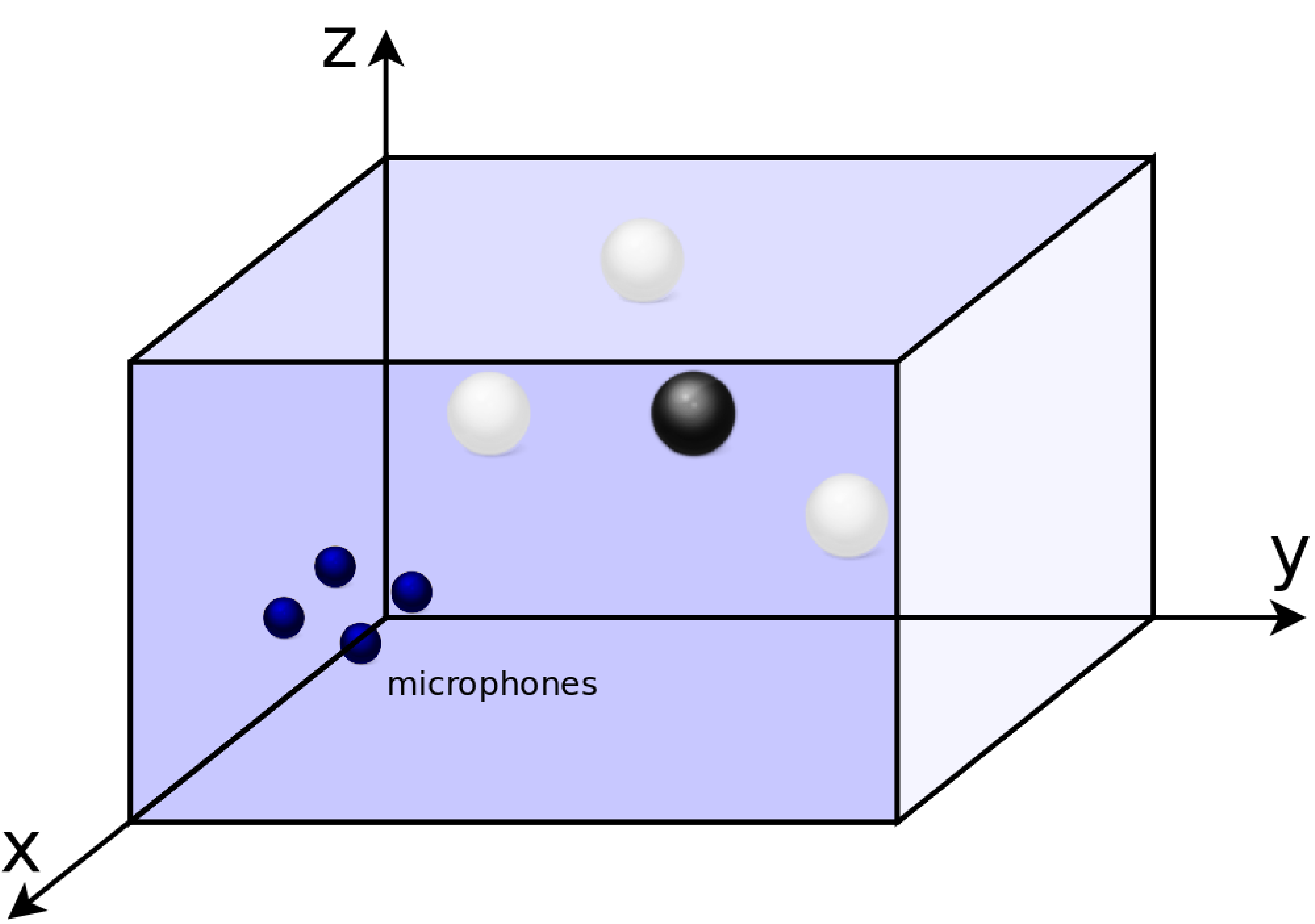}
   \caption{
    A simulated room: There may be multiple microphones, a single target
    sound source, multiple noise sources in a cuboid-shape room with
    acoustically reflective walls \cite{C_Kim_INTERSPEECH_2017_1}. 
    \label{fig:room_diagram}
  }
  \end{center}
  \vspace{-10mm}
\end{figure}
In this section, we briefly review the structure of the Google
Room Simulator for generating simulated utterances
to train acoustic models for speech recognition systems
\cite{C_Kim_INTERSPEECH_2017_1}. We assume a room of a
rectangular cuboid-shape as shown in Fig. \ref{fig:room_diagram}.
Assuming that all the walls of a room reflect acoustically uniformly,
we use the image method to model the Room Impulse Responses (RIRs)
\cite{J_Allen_JASA_1979, E_A_Lehmann_ASPAA_2007,
s_g_mcgovern_mathworks_file_exchange_2013_00,
s_g_mcgovern_applied_acoustics_2009_00}.
In the image method, a real room is acoustically mirrored
with respect to each wall, which results in grids of virtual rooms.
%
In our work for training the acoustic model for Google Home
\cite{C_Kim_INTERSPEECH_2017_1, B_Li_INTERSPEECH_2017_1}, we consider
$17 \times 17 \times 17 = 4913$ virtual rooms for RIR calculation.
Following the image method and assuming that there are $V$ 
virtual sound sources including one real source, 
the impulse response is calculated using the following equation
\cite{ J_Allen_JASA_1979, E_A_Lehmann_ASPAA_2007}:
\begin{align}
    h[n] = \sum_{v = 0}^{V-1} \frac{r^{g_v}}{d_v}
    \delta \left[n -\left \lceil{\frac{d_v f_s}{c_0}}\right \rceil \right],
      \label{eq:h_n_calculation}
\end{align}
where $v$ is the index of each virtual sound source, and $d_v > 0$ is the distance
from that sound source to the microphone, $0 < r < 1$ is
the reflection coefficient of the wall, $g_v$ is
the number of the reflections
to that sound source, $f_s$ is the sampling rate of the RIR, and $c_0$ is the
speed of sound in the air. We use the value of $f_s = 16,000$ \textit{Hz} and
$c_0 = 343$ \textit{m/s} for \textit{Room Simulator} \cite{B_Li_INTERSPEECH_2017_1}.
For $d_v$, $r$, we use numbers created by a random number generator
following specified distributions for each impulse response \cite{C_Kim_INTERSPEECH_2017_1}.

Assuming that there are $I$ sound sources including one target
source and $J$ microphones, the received signal at microphone
$j$ is given by:
\begin{align}
  y_j[n] =  \sum_{i=0}^{I-1} \alpha_{ij} \left(h_{ij}[n] * x_i[n]\right).
  \label{eq:y_j_def}
\end{align}
Since we used a two-microphones system in
\cite{C_Kim_INTERSPEECH_2017_1, B_Li_INTERSPEECH_2017_1}
$J$ is two, and for the number of noise sources, we used a value from zero
up to three with an average of 1.55. Including the target source, the average
number of $I$ is 2.55 \cite{C_Kim_INTERSPEECH_2017_1}.
%
%
%
%
%
%
%
%
%
%

%
%
%
\subsection{Efficient Room Impulse Response filtering}
When the signal and the room impulse response lengths are $N_x$ and
$N_h$ respectively, the number of multiplications $C_{td}$ required for
calculating \eqref{eq:y_j_def} using the time-domain convolution is given by:
\begin{align}
  C_{td} = I \times J \times N_x \times N_h.
  \label{eq:num_mults_linear_convolution}
\end{align}
In our training set used in \cite{C_Kim_INTERSPEECH_2017_1, B_Li_INTERSPEECH_2017_1},
the average utterance length including \emph{non-speech} portions
marked by the Voice Activity Detector (VAD) is 7.31 \textit{s}.
This corresponds to $N_x$ of 116,991 at 16 \textit{kHz}. The average
length of the RIR in the room simulator used in
\cite{C_Kim_INTERSPEECH_2017_1} is 0.243 \textit{s}, corresponding
to the $N_h$ value of 3893.
For Google Home the number of noise sources in our training
set is 1.55 on average.
Thus $I$ is 2.55 including one target source, and $J$ is two, since
we use a two-microphones system in
\cite{C_Kim_INTERSPEECH_2017_1, B_Li_INTERSPEECH_2017_1}.
Thus, if we directly use the linear convolution, it requires
2.32 billion multiplications per utterance from
\eqref{eq:num_mults_linear_convolution}, which is prohibitively large.
Thus, in the ``\textit{Room Simulator}" in
\cite{C_Kim_INTERSPEECH_2017_1},
we used the frequency domain multiplication using \texttt{Kiss FFT}
\cite{M_Borgerding_kiss_fft_2010}. To avoid time aliasing,
the FFT size $N$ must satisfy $N \ge N_x + M - 1$, where
$M$ is the length of the impulse response.
The $j$-th microphone channel
of the simulated signal $y_j[n]$ in \eqref{eq:y_j_def}
is given by:
\begin{align}
  y_{j} [n] = \sum_{i=0}^{I-1}FFT^{-1} \left \{
      FFT \left \{ h_{ij}[n] \right \} \times
      FFT \left \{ x_i[n] \right \}
      \right \}.  
  \label{eq:fir_fft_convolution}
\end{align}
As shown in \eqref{eq:fir_fft_convolution}, we perform
two FFTs, one IFFT, and one complex element-wise multiplications
between two complex spectra for each convolution term in \eqref{eq:y_j_def}.
Assuming that radix-2 FFTs are employed, each FFT or IFFT requires
$\frac{N}{2} \log_2(N)$ multiplications
\cite{A_V_Oppenheim_PrenticeHall_1999}.
In addition, a single element-wise complex multiplication
between two complex spectra is required for each convolution
term in \eqref{eq:y_j_def}.
For real time-domain signals, we need to perform element-wise
multiplications for the lower half-spectrum for
Discrete Fourier Transform (DFT) indices
$0 \le k \le N / 2$ since the spectrum has the Hermitian symmetry
property \cite{A_V_Oppenheim_PrenticeHall_1999}.
From this discussion, we conclude that the number of real multiplications
$C_{\textit{FFT}}$ for calculating \eqref{eq:fir_fft_convolution} is given by:
\begin{align}
  C_{\textit{FFT}} = I J \left(6 N \log_2(N) + 2 N\right).
  \label{eq:c_fft}
\end{align}
For the average $N_x$ of 116,991 mentioned above,
if we assume that the average $N$ is $2 ^ {17}$,  \eqref{eq:c_fft} 
requires 69.5 million multiplications per utterance on average. 
\textit{Room Simulator} for \cite{C_Kim_INTERSPEECH_2017_1, B_Li_INTERSPEECH_2017_1}
was implemented in C++ using the \texttt{Eigen3} linear algebra library \cite{eigenweb}.
So far we have used the \texttt{Kiss FFT} version of FFT in \texttt{Eigen3}
including acoustic model training for Google Home described in
\cite{C_Kim_INTERSPEECH_2017_1, B_Li_INTERSPEECH_2017_1}.
But, to further speed up frontend computation, we switched from
\texttt{Kiss FFT} in \texttt{Eigen3} to a custom  C++ class
implementation which internally uses real FFT in \texttt{FFTW3}.

A more efficient approach is using the OverLap Add (OLA)
FFT filtering \cite{A_V_Oppenheim_PrenticeHall_1999, 
r_crochiere_ieee_tassp_1980_00}.
With the OLA FFT filtering,
the approximate number of real multiplications is given by:
\begin{align}
  C_{\textit{OLA}} = I J \left( 
          \left \lfloor{\frac{N_x}{N - N_h + 1}}\right \rfloor
          \left(4 N \log_2(N) + 2 N  \right)   \right.  \nonumber \\
           \left.
                   \vphantom{\left \lfloor{\frac{N_x}{N - M + 1}}\right \rfloor}
            + 2 N \log_2(N) \right),
  \label{eq:c_ola}
\end{align}
where $N$ is the FFT size, $N_x$ is the length of the entire
signal, and $N_h$ is the length of the impulse response. The term
$4 N \log_2(N) + 2 N $ appears in \eqref{eq:c_ola}, since there is
one FFT, one IFFT, and one element-wise complex multiplication
for $0 \le k \le N / 2$ for each block.
As before, we assumed that each FFT or IFFT requires
$\frac{N}{2} \log_2(N)$ multiplications and one complex multiplication
requires four real multiplications in \eqref{eq:c_ola}.
$\left \lfloor{\frac{N_x}{N - M + 1}}\right \rfloor$
is the number of blocks to process an utterance of length $N_x$.
The $2 N \log_2(N)$ term in \eqref{eq:c_ola} is required for
FFT of the impulse response $h_{ij}[n]$.
For the impulse response $h_{ij}[n]$, we do not
need to repeat it for every block.
In our training set consisting of 22 million utterances mentioned above,
the average $N$ is 116,991 and the average $N_h$ is 3,893.
For this average $N$ and $N_h$, the minimum value of $C_{\textit{OLA}}$
in \eqref{eq:c_ola} is 50.8 millions of real multiplications when $N = 2^{14}$.
The optimal value of $N$ minimizing $C_{\textit{OLA}}$ is different for
different $N_h$ and $N_x$ values. For each filtering, we minimize
$C_{\textit{OLA}}$ by evaluating this equation with different values of
$N = 2^{m}$ where $m$ is an integer.
\subsection{Room Impulse Response Length Selection}
\begin{figure}
  \begin{center}
    \subfloat[] {
      \includegraphics[width=70mm]{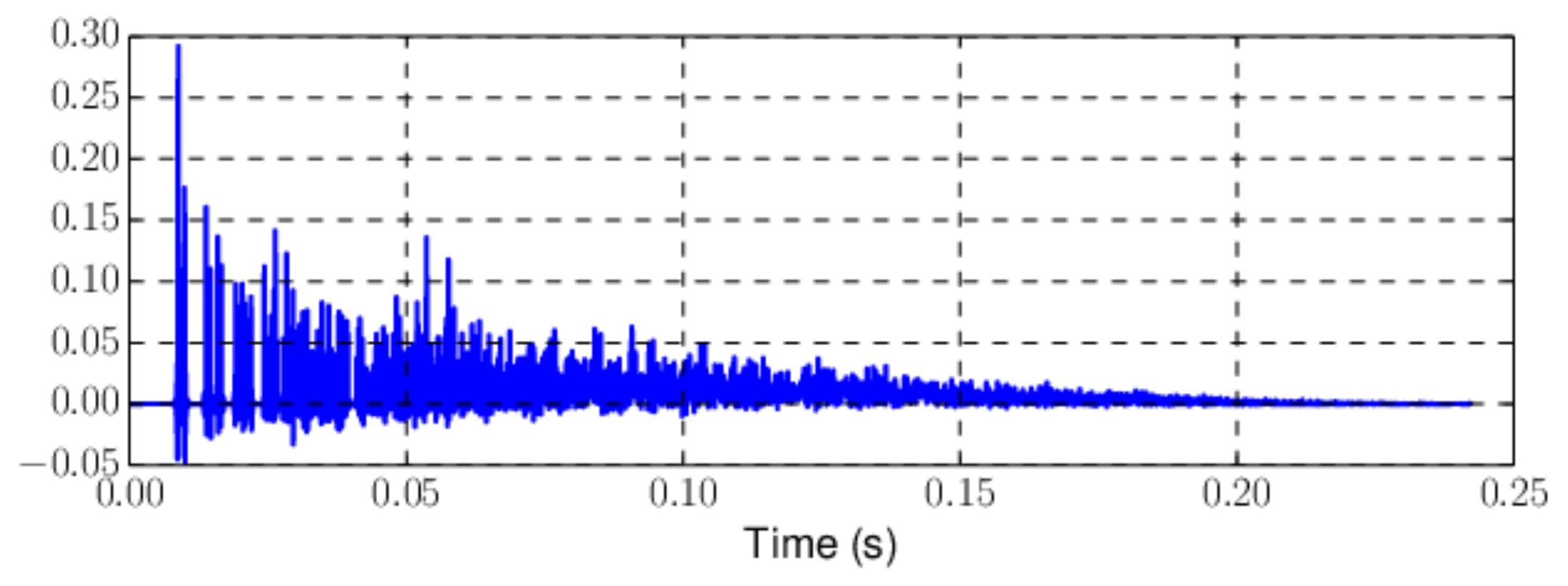}
      \label{fig:rir_cutoff_simulated_sets}
    }
    \vspace{-2mm}
    \\
    \subfloat[] {
      \includegraphics[width=70mm]{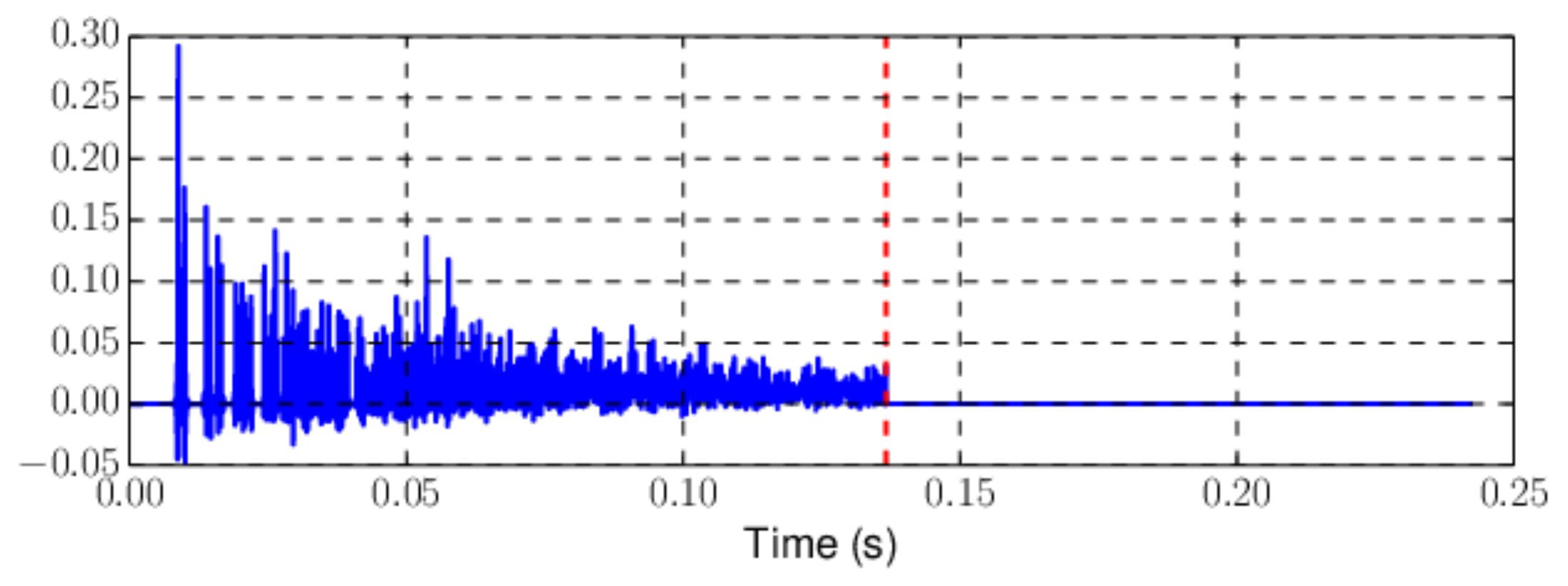}
      \label{fig:rir_cutoff_rerecorded_sets}
    }
    \vspace{-2mm}
    \\
    \caption {
      The simulated room impulse responses generated in the
      ``Room Simulator" described in Sec. \ref{sec:implementation}:
      (a) The impulse response without the RIR tail cut-off,
      (b) with the RIR tail cut-off at 20 \textit{dB}.
    }
   \vspace{-8mm}
  \label{fig:rir_cutoff}
  \end{center}
\end{figure}

%
%
The average length of the Room Impulse Response in the original room
simulator is estimated to be 3893 samples, which corresponds to 0.243
\textit{s}.
We perform a simple RIR tail-cutoff by finding the RIR power threshold
which is $\eta$ \textit{dB} below the maximum power of the RIR
$h_{ij}[n]$:
\begin{align}
  p_{th} = \max \left \{h_{ij}^2[n] \right \} \times 10^{-\frac{\eta}{10}}.
  \label{eq:p_th_def}
\end{align}
Using $p_{th}$, we find the cut-off index $n_c$ which is the smallest
sample index beyond which all the trailing $h_{ij}[n]$ has power
below the threshold $p_{th}$:
\begin{align}
  n_c = \min_{m} \left\{
  m \middle| \max_{n > m} \left\{  h^2[n] \right \} <  p_{th}  \right \}.
\end{align}
Then the final RIR cut-off $\widehat{h}_{ij}[n]$ is given
by the following equation:
\begin{align}
  \widehat{h}_{ij}[n] = h_{ij}[n], \qquad 0 \le n \le n_c + 1.
\end{align}
Fig. \ref{fig:rir_cutoff} shows the original impulse responses and
the corresponding RIR when the cutoff threshold of $\eta$ is used.
To reflect the typical case in
our training set, we used the average room dimension, average
microphone-to-target distance, and average $T_{60}$ value among
3 million room configurations described in \cite{C_Kim_INTERSPEECH_2017_1}.

\begin{figure}
  \begin{center}
    \includegraphics[width=70mm]{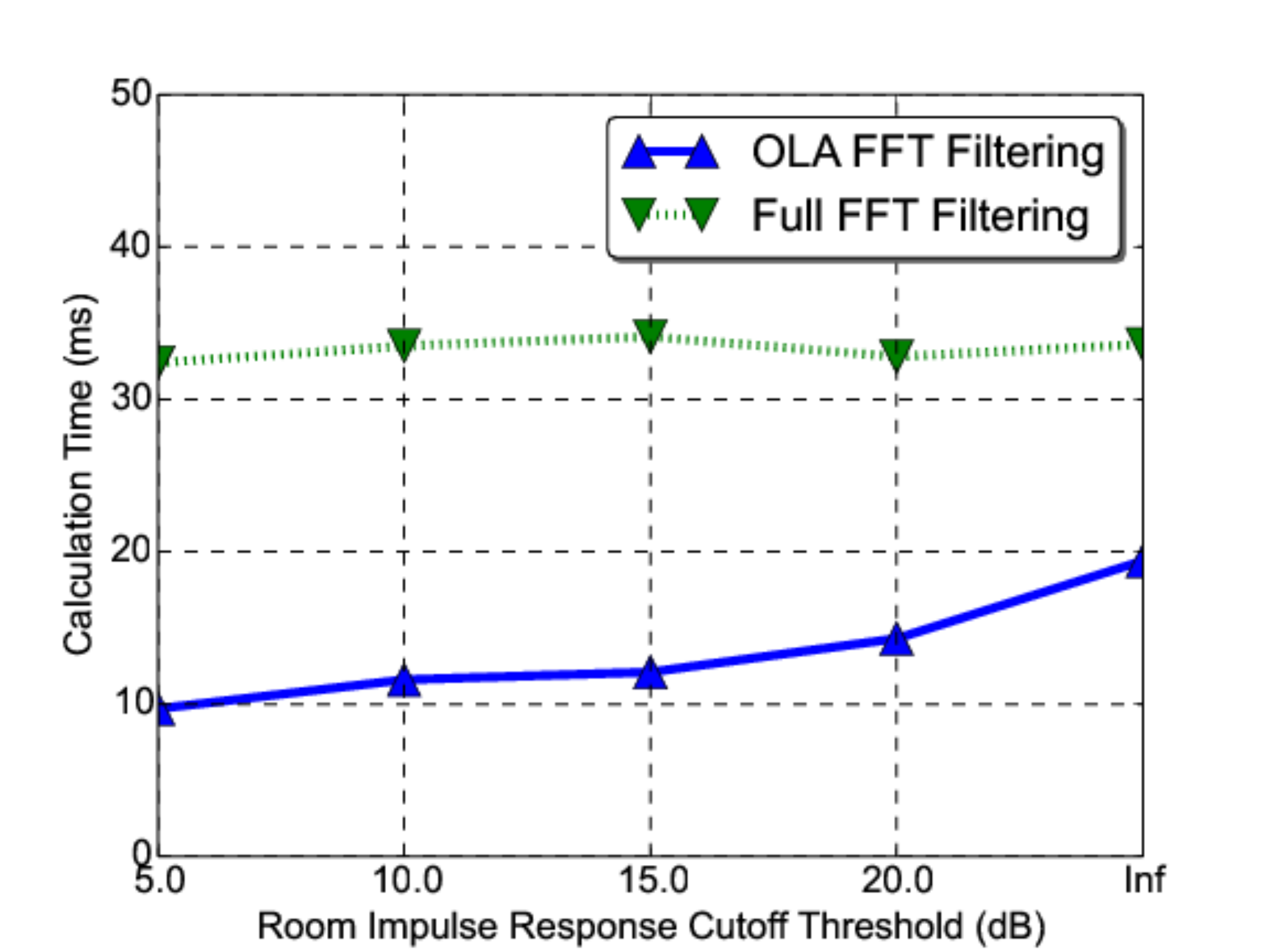}
    \caption {
      \label{fig:rir_cutoff_local_profiling}
      Profiling result on local desktop machine with
      different RIR cut-off threshold with and without
      OverLap Addition FIR filtering.
    }
  \end{center}
  \vspace{-9mm}
\end{figure}

As shown in Table \ref{tbl:local_machine_prof} and \ref{tbl:cpu_percentage},
the RIR tail cut-off at 20 \textit{dB} shows relatively 35.6 \% to 69.4 \%
speed improvement on a local desktop machine and on \texttt{Google Borg}
cluster \cite{A_Verma_eurosys_2015_1}.
Fig. \ref{fig:rir_cutoff_local_profiling} shows the profiling results
on a local machine with different RIR cutoff thresholds $\eta$.
The local machine we used has a single \texttt{Intel(R) Xeon(R) E5-1650 @ 3.20GHz}
CPU with 6 cores and 32 GB of memory. In Fig. \ref{fig:rir_cutoff_local_profiling},
we observe that the computational cost becomes less for the OLA filtering
case as we cut the tail portion of the RIR more. For the full FFT case,
theoretically, it should remain almost constant regardless of the
impulse response length, but due to variation
in profiling measurement, there is some small variation.
%
%
%
%
%
%
%
\begin{table}
  \renewcommand{\arraystretch}{1.3}
  \centering
  \caption{\label{tbl:local_machine_prof}The profiling result for processing a single utterance using
           the ``Room Simulator" on a local desktop machine}
\begin{tabular}{| c | c |}
  \hline
                         & \specialcell{Avg. Time per \\ Utterance (\textit{ms})} \\
   \hline
   \hline
  \specialcell{Original \textit{Room Simulator} in \cite{C_Kim_INTERSPEECH_2017_1}}                            &  320.8 \textit{ms} \\
   \hline
  \specialcell{\texttt{FFTW3} Real FFT Filtering }                  &  44.2  \textit{ms} \\
   \hline
  \specialcell{+OLA filtering }                             &  19.4  \textit{ms} \\
   \hline
  \specialcell{+20 dB RIR cut-off}        &  14.3  \textit{ms} \\
  \hline  
\end{tabular}
\end{table}
\begin{table*}
  \caption{\label{tbl:cpu_percentage} {The CPU usage portion of the FFT
          and \textit{Room Simulator} with the respect to the entire
          CPU pipeline in Fig. \ref{fig:gpu_cpu_structure}  and \\ relative
          speed up measured in terms of execution time with respect to our baseline 
          in \cite{C_Kim_INTERSPEECH_2017_1} on \texttt{Google Borg} 
          cluster \cite{A_Verma_eurosys_2015_1}.}}
  \vspace{2mm}
  \centerline{
    \begin{tabular}{| c | c | c | c | c | c | c |}
      \hline
            &  \specialcell{Original system in \cite{C_Kim_INTERSPEECH_2017_1}}
            &  \specialcell{\texttt{FFTW3} OLA Filter   }
            &  \specialcell{\texttt{FFTW3} OLA Filter \\ with 20 dB RIR  Cut-off}
            &  \specialcell{\texttt{FFTW3} OLA Filter \\ with 10 dB RIR  Cut-off}  \\
      \hline \hline
      FFT portion (\%)       & 79.27 \%  &  10.11 \% &  \textbf{6.23} \% & 5.08 \% \\
      \hline
      \specialcell{Relative Speed Up \\ in FFT Portion (\%) }
                             &  -    & 34.0 times    &  \textbf{57.6 times} & 59.6 times       \\
      \hline \hline
      \specialcell{Entire Acoustic \\ Simulation (\%) }
                             & 80.01 \%  &  13.30 \% &  \textbf{9.69} \% & 8.05 \%  \\
      \hline
      \specialcell{Relative Speed Up \\ in Acoustic Simulation Portion (\%) }
                             &  -    &  26.1 times    &  \textbf{37.3 times}  & 45.7 times       \\
      \hline
    \end{tabular}
  }
\end{table*}
\begin{table*}
  \caption{\label{tbl:results} {Speech recognition experimental result in terms of Word Error Rates (WERs) \\
          with and without MTR using the room simulation system in \cite{C_Kim_INTERSPEECH_2017_1} and different RIR cutoff thresholds.}}
  \vspace{2mm}
  \centerline{
    \begin{tabular}{| c | c | c | c | c | c | c |}
      \hline
            &  \specialcell{Baseline \\ without  MTR}
            &  \specialcell{MTR using  \\ One-time Batch \\ Room Simulation}
            &  \specialcell{Baseline   \\ (On-the-fly    \\ Room Simulation)}
            &  \specialcell{20 dB RIR  \\ Cut-off}
            &  \specialcell{10 dB RIR  \\ Cut-off}
            &  \specialcell{5 dB RIR   \\ Cut-off}  \\
      \hline \hline
      Original Test Set        & 11.70 \%   &  12.07 \% & 11.95 \%  & \textbf{11.43} \%  & 10.75 \% &  11.20  \%  \\
      \hline
      \textit{Simulated Noisy Set A}    & 20.75 \%  &  15.58 \% & 14.88 \%  & \textbf{13.96} \% & 13.59 \%  &  14.99 \%  \\
      \hline
      \textit{Simulated Noisy Set B}    & 50.78 \%  &  22.47 \% & 20.64 \%  & \textbf{19.58} \% & 21.37 \%  &   28.27 \%  \\
      \hline
      \specialcell{\textit{Device 1}} & 52.56 \%  &  22.39 \% & 21.69 \%  & \textbf{21.35} \% & 23.20 \%   &  29.53 \%  \\
      \hline
      \specialcell{\textit{Device 2}} & 51.59 \%  &  22.12 \% & 21.62 \%  &  \textbf{21.26} \% & 22.90 \%   & 29.39 \%  \\
      \hline
      \specialcell{\textit{Device 3}} & 54.89 \%  &  23.42 \% & 22.29 \%  &  \textbf{22.86} \%  & 26.47 \%  &  36.71 \%  \\
      \hline
      \specialcell{\textit{Device 3} \\ (Noisy Condition)  }
                              &  72.09 \%  &  36.21 \% & 35.88 \%  &  \textbf{35.83} \%  & 39.99 \%  &  51.11 \%  \\
      \hline
      \specialcell{\textit{Device 3} \\ (Multi-Talker Condition)  }
                             &  74.60 \%   &  47.63 \% & 46.03 \%  &  \textbf{46.21} \%  & 48.36 \%  & 58.31 \%   \\
      \hline
    \end{tabular}
  }
\end{table*}
%
%
%
%
 
%
%
%
%
\section{Experimental results}
\label{sec:experimental_results}
%
%
%
%
%

In this section, we present speech recognition results and CPU profiling
results obtained using \textit{Room Simulator} with different RIR cut-off
thresholds. The acoustic modeling structure to obtain
 experimental results in this section is somewhat different
from those described in \cite{C_Kim_INTERSPEECH_2017_1, B_Li_INTERSPEECH_2017_1}
  for faster training.
We use a single-channel 128 log-Mel feature whose window
size is 32 \textit{ms}. The interval between successive frame is 10
\textit{ms}. The low and upper cutoff frequencies of the Mel filterbank
are 125 \textit{Hz} and 7500 \textit{Hz} respectively.
Since it has been shown that long-duration features represented by overlapping
features are helpful \cite{H_Sak_INTERSPEECH_2015_1}, four frames
are stacked together.
Thus we use a context dependent feature consisting of 512 elements
given by 128 (the size of the log-mel feature) x 4 (number of stacked frames).
  This input is down-sampled by a factor of three 
  \cite{g_pundak_interspeech_2016_00}.
  The feature is processed by a typical multi-layer Long Short-Term Memory 
  (LSTM) \cite{S_Hochreiter_neural_computation_1997_00, T_Sainath_ICASSP_2015_1} acoustic model.
We use 5-layer LSTMs with 768 units in each layer.
The output of the final LSTM layer is passed to a softmax layer.
The softmax layer has 8192 nodes corresponding to the number of tied
context-dependent phones in our ASR system. The output state label is delayed
by five frames, since it was observed that
the information about future frames improves the prediction of the current frame
  \cite{H_Sak_INTERSPEECH_2014_1, M_Schuster_ieee_trans_signal_processing_1997}.
The acoustic model was trained using the Cross-Entropy (CE) loss
as the objective function, using precomputed alignments for utterance as targets.
To obtain results in Table \ref{tbl:results}, we trained for about 45 epochs.
\begin{table}
  \renewcommand{\arraystretch}{1.3}
  \centering
  \caption{\label{tbl:avg_t_60}Average $T_{60}$ Time of the Simulated 
  Training Set and Simulated Test Sets A and B.}
\begin{tabular}{| c | c | c | c |}
  \hline
                         & \specialcell{Simulated \\ Training Set}
                         & \specialcell{Simulated \\ Noisy Set A}
                         & \specialcell{Simulated \\ Noisy Set B}  \\
   \hline
  \specialcell{Average $T_{60}$ \textit{(s)}}
      & 0.482 \textit{s}
      & 0.167 \textit{s}
      & 0.479 \textit{s}\\
  \hline
\end{tabular}
\vspace{-7mm}
\end{table}
For training, we used an anonymized and hand-transcribed
22-million English utterances (18,000-hr) set. The training set is the same as what we
used in \cite{C_Kim_INTERSPEECH_2017_1, B_Li_INTERSPEECH_2017_1}.
For evaluation, we used around 15-hour of utterances (13,795 utterances)
obtained from anonymized mobile voice search data. 
  We also generate noisy evaluation
sets from this relatively clean voice search data.
We use both simulated and rerecorded noisy sets.
The average reverberation time in $T_{60}$ of the simulated training set
and two simulated test sets are shown in Table \ref{tbl:avg_t_60}.
These two simulated test sets are named \textit{Simulated Noisy Set A} and 
\textit{Simulated Noisy Set B} respectively.

Since our objective is deploying our speech recognition
systems on far-field standalone devices such as Google Home,
we rerecorded these evaluation sets using the actual hardware
in far-field environment.
Note that the actual Google Home hardware has two microphones
with microphone spacing of 7.1 cm. In our experiments in this section,
we selected the first channel out of two channel data.
 Three different devices
were used in rerecording, and each device was placed in five
different locations in an actual room resembling a real living
room. These devices are listed in Table \ref{tbl:results} as ``Device 1'', ``Device 2'',
and ``Device 3''. As shown in Table \ref{tbl:results},
we observe that the RIR cut-off up to 20 \textit{dB} threshold does not
adversely affect the performance. However, if we cut the
RIR to 5 \textit{dB} threshold, then the performance under far-field
environment becomes significantly worse. This observation also confirms
that far-field speech recognition benefits from
RIRs with sufficiently long tails in the training set.
Table \ref{tbl:cpu_percentage} shows how much CPU resource was used
when training is done on \texttt{Google Borg} cluster
\cite{A_Verma_eurosys_2015_1} using the CPU/GPU
training architecture in Fig. \ref{fig:gpu_cpu_structure}. We observe that
if we use the \texttt{FFTW3}-based OLA filter using 20 \textit{dB} RIR cutoff, we
may obtain 57.6 times speed-up in the FFT portion. The entire speed-up
of \textit{Room Simulator} portion is 37.3 times.
\section{Conclusions}
In this paper, we describe how to efficiently implement
an acoustic room simulator to generate large-scale
simulated data for training deep neural networks.
We implement an efficient OverLap
Addition (OLA) based filtering using the open-source \texttt{FFTW3}
library. We investigate into the effects of the Room
Impulse Response (RIR) lengths. We conclude that we can cut
the tail portions of RIRs whose power is less than 20 \textit{dB}
below the maximum power without sacrificing speech recognition accuracy.
However, if we cut off RIR more than that, we observe it adversely
affects the performance for reverberant cases.
Using the approaches mentioned here, we could reduce the
room simulator portion in the CPU usage down to 9.69 \%
in CPU/GPU training architecture. Profiling result shows that
we obtain 22.4 times speed-up on a local desktop machine and 37.3 times
speed up on \texttt{Google Borg} cluster.
%
%
\eightpt
\bibliographystyle{IEEEtran}
\bibliography{../../common_bib_file/common_bib_file}

\begin{thebibliography}{10}
\providecommand{\url}[1]{#1}
\csname url@samestyle\endcsname
\providecommand{\newblock}{\relax}
\providecommand{\bibinfo}[2]{#2}
\providecommand{\BIBentrySTDinterwordspacing}{\spaceskip=0pt\relax}
\providecommand{\BIBentryALTinterwordstretchfactor}{4}
\providecommand{\BIBentryALTinterwordspacing}{\spaceskip=\fontdimen2\font plus
\BIBentryALTinterwordstretchfactor\fontdimen3\font minus
  \fontdimen4\font\relax}
\providecommand{\BIBforeignlanguage}[2]{{%
\expandafter\ifx\csname l@#1\endcsname\relax
\typeout{** WARNING: IEEEtran.bst: No hyphenation pattern has been}%
\typeout{** loaded for the language `#1'. Using the pattern for}%
\typeout{** the default language instead.}%
\else
\language=\csname l@#1\endcsname
\fi
#2}}
\providecommand{\BIBdecl}{\relax}
\BIBdecl

\bibitem{C_Kim_INTERSPEECH_2017_1}
{{\chanwcom}, A. Misra, K.K. Chin, T. Hughes, A. Narayanan, T. Sainath, and M.
  Bacchiani}, ``{Generation of simulated utterances in virtual rooms to train
  deep-neural networks for far-field speech recognition in Google Home},'' in
  \emph{INTERSPEECH-2017}, Aug. 2017, pp. {379--383}.

\bibitem{E_Variani_INTERSPEECH_2017_01}
\BIBentryALTinterwordspacing
{E. Variani, T. Bagby, E. McDermott, and M. Bacchiani}, ``{End-to-end training
  of acoustic models for large vocabulary continuous speech recognition with
  tensorflow},'' in \emph{{INTERSPEECH-2017}}, 2017, pp. 1641--1645. [Online].
  Available: \url{http://dx.doi.org/10.21437/Interspeech.2017-1284}
\BIBentrySTDinterwordspacing

\bibitem{Seltzer2013DNNAurora4}
{M. Seltzer, D. Yu, and Y.-Q. Wang}, ``An investigation of deep neural networks
  for noise robust speech recognition,'' in \emph{Int. Conf. Acoust. Speech,
  and Signal Processing}, 2013, pp. 7398--7402.

\bibitem{Yu2013FeatureLearningDNN}
{D. Yu, M. L. Seltzer, J. Li, J.-T. Huang, and F. Seide}, ``Feature learning in
  deep neural networks - studies on speech recognition tasks,'' in
  \emph{Proceedings of the International Conference on Learning
  Representations}, 2013.

\bibitem{V_Vanhoucke_Deep_Learning_NIPS_Workshop_2011}
{V. Vanhoucke, A. Senior, and M. Z. Mao}, ``{Improving the speed of neural
  networks on CPUs},'' in \emph{Deep Learning and Unsupervised Feature Learning
  NIPS Workshop}, 2011.

\bibitem{G_Hinton_IEEE_Signal_Process_Mag_2012}
{G. Hinton, L. Deng, D. Yu, G. E. Dahl, A. Mohamed, N. Jaitly, A. Senior, V.
  Vanhoucke, P. Nguyen, T. Sainath, and B. Kingsbury}, ``{Deep neural networks
  for acoustic modeling in speech recognition: The shared views of four
  research groups},'' \emph{IEEE Signal Processing Magazine}, vol.~29, no.~6,
  Nov.

\bibitem{T_Sainath_IEEETran_2017_1}
{T. Sainath, R. J. Weiss, K. W. Wilson, B. Li, A. Narayanan, E. Variani, M.
  Bacchiani, I. Shafran, A. Senior, K. Chin, A. Misra, and {\chanwcom}},
  ``{Multichannel signal processing with deep neural networks for automatic
  speech recognition},'' \emph{IEEE/ACM Trans. Audio, Speech, Lang. Process.},
  Feb. 2017.

\bibitem{T_Sainath_Book_Chapter_2017_1}
------, ``{Raw Multichannel Processing Using Deep Neural Networks},'' in
  \emph{{New Era for Robust Speech Recognition: Exploiting Deep Learning}}, {S.
  Watanabe, M. Delcroix, F. Metze, and J. R. Hershey}, Ed.\hskip 1em plus 0.5em
  minus 0.4em\relax Springer, Oct. 2017.

\bibitem{B_Li_INTERSPEECH_2017_1}
{{B. Li, T. Sainath, A. Narayanan, J. Caroselli, M. Bacchiani, A. Misra, I.
  Shafran, H. Sak, G. Pundak, K. Chin, K-C Sim, R. Weiss, K. Wilson, E.
  Variani, {\chanwcom}, O. Siohan, M. Weintraub, E. McDermott, R. Rose, and M.
  Shannon}}, ``{Acoustic modeling for Google Home},'' in
  \emph{INTERSPEECH-2017}, Aug. 2017, pp. {399--403}.

\bibitem{C_Kim_IEEETran_2016_1}
{\chanwcom} and R.~M. Stern, ``{Power-Normalized Cepstral Coefficients (PNCC)
  for Robust Speech Recognition},'' \emph{IEEE/ACM Trans. Audio, Speech, Lang.
  Process.}, pp. 1315--1329, July 2016.

\bibitem{U_H_Yapanel_SpeechComm_2008}
{U. H. Yapanel and J. H. L. Hansen}, ``{A new perceptually motivated MVDR-based
  acoustic front-end (PMVDR) for robust automatic speech recognition},''
  \emph{{Speech Communication}}, vol.~50, no.~2, pp. 142--152, {Feb.} 2008.

\bibitem{C_Kim_ICASSP_2010_1}
{{\chanwcom} and R. M. Stern}, ``Feature extraction for robust speech
  recognition based on maximizing the sharpness of the power distribution and
  on power flooring,'' in \emph{IEEE Int. Conf. on Acoustics, Speech, and
  Signal Processing}, March 2010, pp. 4574--4577.

\bibitem{C_Kim_INTERSPEECH_2010_2}
------, ``Nonlinear enhancement of onset for robust speech recognition,'' in
  \emph{INTERSPEECH-2010}, Sept. 2010, pp. 2058--2061.

\bibitem{C_Kim_INTERSPEECH_2014_2}
{{\chanwcom}, K. Chin, M. Bacchiani, and R. M. Stern}, ``Robust speech
  recognition using temporal masking and thresholding algorithm,'' in
  \emph{INTERSPEECH-2014}, Sept. 2014, pp. 2734--2738.

\bibitem{T_Nekatani_ICASSP_2017_1}
{T. Nakatani, N. Ito, T. Higuchi, S. Araki, and K. Kinoshita}, ``{Integrating
  DNN-based and spatial clustering-based mask estimation for robust MVDR
  beamforming},'' in \emph{IEEE Int. Conf. Acoust., Speech, Signal Processing},
  March 2017, pp. 286--290.

\bibitem{T_Higuchi_ICASSP_2016_1}
{T. Higuchi and N. Ito and T. Yoshioka and T. Nakatani}, ``{Robust MVDR
  beamforming using time-frequency masks for online/offline ASR in noise},'' in
  \emph{IEEE Int. Conf. Acoust., Speech, Signal Processing}, March 2016, pp.
  5210--5214.

\bibitem{H_Erdogan_INTERSPEECH_2016_1}
{H. Erdogan, J. R. Hershey, S. Watanabe, M. Mandel, J. Roux}, ``{Improved MVDR
  Beamforming Using Single-Channel Mask Prediction Networks},'' in
  \emph{INTERSPEECH-2016}, Sept 2016, pp. 1981--1985.

\bibitem{C_Kim_INTERSPEECH_2015_1}
{{\chanwcom} and K. K. Chin}, ``Sound source separation algorithm using phase
  difference and angle distribution modeling near the target,'' in
  \emph{INTERSPEECH-2015}, Sept. 2015, pp. 751--755.

\bibitem{C_Kim_INTERSPEECH_2009_1}
{{\chanwcom}, K. Kumar, B. Raj, and R. M. Stern}, ``Signal separation for
  robust speech recognition based on phase difference information obtained in
  the frequency domain,'' in \emph{INTERSPEECH-2009}, {Sept.} 2009, pp.
  2495--2498.

\bibitem{C_Kim_ICASSP_2012_2}
{{\chanwcom}, C. Khawand, and R. M. Stern}, ``Two-microphone source separation
  algorithm based on statistical modeling of angle distributions,'' in
  \emph{IEEE Int. Conf. on Acoustics, Speech, and Signal Processing}, March
  2012, pp. 4629--4632.

\bibitem{C_Kim_ICASSP_2011_2}
{{\chanwcom}, K. Kumar, and R. M. Stern}, ``Binaural sound source separation
  motivated by auditory processing,'' in \emph{IEEE Int. Conf. on Acoustics,
  Speech, and Signal Processing}, May 2011, pp. 5072--5075.

\bibitem{J_Allen_JASA_1979}
J.~Allen and D.~Berkley, ``Image method for efficiently simulating small-room
  acoustics,'' \emph{J. Acoust. Soc. Am.}, vol.~65, no.~4, pp. 943--950, April
  1979.

\bibitem{E_A_Lehmann_ASPAA_2007}
{E. A. Lehmann, A. M. Johansson, and S. Nordholm}, ``Reverberation-time
  prediction method for room impulse responses simulated with the image-source
  model,'' in \emph{2007 IEEE Workshop on Applications of Signal Processing to
  Audio and Acoustics}, Oct. 2007, pp. 159--162.

\bibitem{s_g_mcgovern_mathworks_file_exchange_2013_00}
\BIBentryALTinterwordspacing
{S. G. McGovern}. {room impulse response generator}. [Online]. Available:
  \url{{https://www.mathworks.com/matlabcentral/fileexchange/5116-room-impulse-response-generator}}
\BIBentrySTDinterwordspacing

\bibitem{s_g_mcgovern_applied_acoustics_2009_00}
------, ``Fast image method for impulse response calculations of box-shaped
  rooms,'' \emph{Applied Acoustics}, vol.~70, no.~1, pp. 182 -- 189, 2009.

\bibitem{M_Borgerding_kiss_fft_2010}
{M. Borgerding}, ``Kiss fft version 1.2.9,''
  https://sourceforge.net/projects/kissfft/, 2010.

\bibitem{A_V_Oppenheim_PrenticeHall_1999}
{A. V. Oppenheim and R. W.Scafer, with .J. R. Buck}, \emph{Discrete-time Signal
  Processing}, 2nd~ed.\hskip 1em plus 0.5em minus 0.4em\relax Englewood-Cliffs,
  NJ: Prentice-Hall, 1998.

\bibitem{eigenweb}
{G. Guennebaud, B. Jacob et al.}, ``Eigen v3,'' http://eigen.tuxfamily.org,
  2010.

\bibitem{r_crochiere_ieee_tassp_1980_00}
R.~Crochiere, ``{A weighted overlap-add method of short-time Fourier
  analysis/synthesis},'' \emph{IEEE Trans. Acoust., Speech, and Signal
  Processing}, vol.~28, no.~1, pp. 99--102, feb 1980.

\bibitem{A_Verma_eurosys_2015_1}
A.~Verma, L.~Pedrosa, M.~R. Korupolu, D.~Oppenheimer, E.~Tune, and J.~Wilkes,
  ``Large-scale cluster management at {Google} with {Borg},'' in
  \emph{Proceedings of the European Conference on Computer Systems (EuroSys)},
  Bordeaux, France, 2015.

\bibitem{H_Sak_INTERSPEECH_2015_1}
{H. Sak, A. Senior, K. Rao, and F. Beaufays}, ``{Fast and Accurate Recurrent
  Neural Network Acoustic Models for Speech Recognition},'' in
  \emph{{INTERSPEECH-2015}}, Sept. 2015, pp. 1468--1472.

\bibitem{g_pundak_interspeech_2016_00}
\BIBentryALTinterwordspacing
{G. Pundak and T. N. Sainath}, ``{Lower Frame Rate Neural Network Acoustic
  Models},'' 2016, pp. 22--26. [Online]. Available:
  \url{http://dx.doi.org/10.21437/Interspeech.2016-275}
\BIBentrySTDinterwordspacing

\bibitem{S_Hochreiter_neural_computation_1997_00}
{S. Hochreiter and J{\"u}rgen Schmidhuber}, ``{Long Short-term Memory},''
  \emph{Neural Computation}, no.~9, pp. 1735--1780, Nov. 1997.

\bibitem{T_Sainath_ICASSP_2015_1}
{T. N. Sainath, O. Vinyals, A. Senior, and H. Sak}, ``{Convolutional, long
  short-term memory, fully connected deep neural networks},'' in \emph{IEEE
  Int. Conf. Acoust., Speech and Signal Processing}, Apr. 2015, pp. 4580--4584.

\bibitem{H_Sak_INTERSPEECH_2014_1}
{H. Sak, A. Senior, and F. Beaufays}, ``{Long short-term memory recurrent
  neural network architectures for large scale acoustic modeling},'' in
  \emph{{INTERSPEECH-2014}}, Sept. 2014, pp. 338--342.

\bibitem{M_Schuster_ieee_trans_signal_processing_1997}
M.~Schuster and K.~Paliwal, ``Bidirectional recurrent neural networks,''
  \emph{IEEE Transactions on Signal Processing}, vol.~45, no.~11, pp.
  2673--2681, 1997.

\end{thebibliography}
%
%
\end{document}